\documentclass[pre,aps,preprint,showpacs]{revtex4}
\usepackage{revsymb,latexsym,amssymb,amsmath,comment}
\usepackage{graphicx,psfrag,subfigure,stmaryrd}

\begin{document}
\author{Steffen Trimper}
\affiliation{Fachbereich Physik,
Martin-Luther-Universit\"at, D-06099 Halle Germany}
\email{steffen.trimper@physik.uni-halle.de}
\title{Master Equation and Two Heat Reservoirs}
\date{\today }
\begin{abstract}
We analyze a simple spin-flip process under the presence of two heat reservoirs. While one flip process 
is triggered by a bath at temperature $T$, the inverse process is activated by a bath at 
a different temperature $T\,^{\prime}$. The situation can be described by using a master equation 
approach in a second quantized Hamiltonian formulation. The stationary solution leads to a generalized 
Fermi-Dirac distribution with an effective temperature $T_e$. Likewise the relaxation time is given in terms  
of $T_e$. Introducing a spin-representation we perform a Landau expansion for the averaged spin 
$\langle \sigma \rangle$ as order parameter and consequently, a free energy functional can 
be derived. Owing to the two reservoirs the model is invariant with respect to a simultaneous 
change  $\sigma \leftrightarrow -\, \sigma $ and $ T \leftrightarrow T\,^{\prime}$. This new symmetry 
generates a third order term in the free energy which gives rise a dynamically induced first 
order transition. 


\end{abstract}

\maketitle

\section{Introduction}

\noindent Whereas equilibrium statistical mechanics has based on a secure theoretical foundation, 
this is far from being the case in nonequilibrium. The feature of equilibrium phenomena is the existence of a 
probability distribution describing the statistical properties of these systems. In general this 
distribution depends on the interaction among the particles and the temperature of a single external 
source called a heat bath. In nonequilibrium the situation is different and consequently also the 
methods in attacking the problems are different. A variety of processes are described by Markovian 
models, where the dynamical process depends only on the present configuration of the system. 
The master equation is one important tool for describing different stochastic processes on a complex 
energy landscape \cite{gard}. The inputs, required for the master equation \cite{vk}, are a set of states 
and a set of transition rates between those states, for a very recent approach see \cite{sun}. 
Often the transition rates are determined in 
according to the principle of detailed balance, in particular in case the system is coupled to a 
single heat bath with a certain but fixed temperature. Thus for stochastic jump processes the rates 
are assumed to follow an Arrhenius ansatz with an activation energy in terms of the temperature of 
the underlying heat bath. Otherwise there is no necessity for having only one bath. Therefore we 
consider here a simple model with two separate heat baths with different temperatures. To be specific 
let us study a annihilation and creation process of particles or an equivalent spin-flip process. 
However both processes should be activated by different heat baths. While the spin-flip up-down 
is triggered by a bath at the temperate $T$, the reversed down-up-flip is activated by the heat bath 
at the different temperature $T^{\prime}$. Apparently both flip rates are likewise determined by 
different temperatures. One could speculate about a generalization by introducing as many heat 
baths as energy levels exit, i.e. each state is related to its own bath and maybe there is an flow between 
the baths which established a typical nonequilibrium situation.\\ 
In the present paper we are interested in a two-level model which can be visualized in terms of a 
spin variable or alternatively by lattice gas variables. The flip process will be organized by 
a coupling to two local baths. An appropriate method to study such a situation is given by the master 
equation approach formulated in terms of second quantized operators \cite{doi,gra,sp,satr,sti,mg,gs}. 
In that approach the flip-processes are described by creation and annihilation operators, whereas the 
temperate dependence of the rates are incorporated in the approach by using a Heisenberg-like picture 
\cite{sctr,st1,mts}. The approach is generalized in such a manner which enables us to consider 
two different heat baths.\\
The analysis can be grouped into the current interest in studying systems with different heat reservoirs 
\cite{cglv,tag,br,e,s,vrmh,sb,vs,rws,pel,s1,ta,raf,b1,zb,kn,ah}. The analysis is motivated by 
searching for some generic features of nonequilibrium steady states. In particular, the question 
arises for a universal behavior under nonequilibrium conditions. As an example, a two-temperature, 
kinetic Ising model is investigated in \cite{cglv} and extended to a diffusive kinetic system in \cite{br}.
The authors found a bicritical point, where two nonequilibrium critical lines meet. The analysis is strongly 
supported by Monte Carlo simulations in two dimensions. A similar simulation has been performed 
studying a two-temperature lattice gas model with repulsive interactions \cite{s}. 
The two dimensional nonequilibrium Ising model with competing dynamics induced by two heat baths had been 
studied in \cite{tag}. Despite of the two reservoirs the critical exponents belong to the same universality class 
as the corresponding equilibrium model. Alternatively a two temperature lattice gas model with repulsive 
interactions is studied \cite{s,s1}. Hereby the nonequilibrium transition remains continuous unlike in our 
approach. Another field of interest is the Carnot engine, Carnot refrigerator \cite{vrmh,b1} including a 
thermally driven ratchet under periodic dichotomous temperature change \cite{cglv}, which can be likewise  
characterized by two reservoirs. General aspects of a thermodynamic cycle with open flow had been considered in 
\cite{rws} and a rectification of the Clausius inequality is recently discussed in \cite{ah}. 
Totally different physical situations occur, when the flow of complex fluids had been 
analyzed under different heat sources \cite{e}, or in case of a nonlinear oscillator coupled to various 
heat baths \cite{vs}. As pointed out in \cite{pel} magnetic systems with annealed degrees of freedom 
are predestined to offer some features of a two-temperature systems. An interesting physical 
explanation for a two reservoir system is discussed recently \cite{raf}, where the fast and the 
slow variables of a Hamiltonian system are related to different heat baths. In \cite{raf} it is 
demonstrated that the Onsager relations do not apply if the two baths are not too close. Apparently 
transport properties are determined by the heat sources. In \cite{zb} the occurrence of anomalous heat 
conductivity in a one-dimensional non-Markov process is studied, whereas in \cite{kn} a hidden heat 
transfer is observed, when the nonequilibrium steady states are maintained by two heat baths. 
Very recently in a series of papers \cite{a} the phase space probability density for steady heat 
flow is discussed. In that case the two reservoirs are mutually connected leading to a flow.\\
As mentioned above we study a spin-flip process under the influence of two heat reservoirs. Because 
this process is on a mesoscopic description related to model A in the classification of Hohenberg and 
Halperin \cite{hh} we also analyze the critical dynamics under a two-temperature reservoir.

\section{Quantum Approach to Nonequilibrium}

\noindent The further analysis is based on a master equation which is written in the form  
\begin{equation}
\partial_tP(\vec n,t)=\mathcal{L}P(\vec n,t)\,. 
\label{ma}
\end{equation}
Here $P(\vec n, t)$ is the joint probability density that a certain configuration,  
characterized by a state vector $\vec n = (n_1, n_2 \dots n_N) $, is realized at time $t$. 
In a lattice gas description each point is either empty or single occupied leading to 
$n_i = 0,\, 1$. Since these numbers can be considered as the eigenvalues of the 
particle number operator and because of the similarity of the evolution equation (\ref{ma}) 
to the Schr\"odinger equation one can introduce a quantum formulation of the master equation. 
This is firstly done by Doi \cite{doi} for a Bose-like system and later by other authors for 
spin-operators \cite{gra,sp,satr}, for reviews compare \cite{mg,sti,gs}. The dynamics of the system 
is determined completely by the the evolution operator $\mathcal{L}$ and the commutation 
relations of the underlying operators. In case of using Pauli-operators the restrictions 
for the occupation numbers to empty and single occupied states is guaranteed, see Eq.~(\ref{com}). 
To transform the basic equation (\ref{ma}) in a second quantized form one has to relate 
the probability distribution $P(\vec n,t)$ to a state vector $\mid F(t)\, \rangle$ 
in a Fock-space according to $P(\vec n,t) = \langle \vec n\mid F(t)\rangle$. If the state 
vectors $\mid \vec n \rangle$ are a complete set then the last relation implies 
the expansion 
\begin{equation}
\mid F(t) \rangle = \sum_{n_i} P(\vec n,t) \mid \vec n \rangle\,.
\label{fo2}
\end{equation}  
Under this transformation Eq.~(\ref{ma}) can be rewritten as an equivalent 
equation in a Fock-space
\begin{equation}
\partial_t \mid F(t)\rangle = L \mid F(t) \rangle\,.
\label{fo1}
\end{equation}
where the operator $L$ is determined in such a manner that its matrix elements correspond to 
$\mathcal{L}$. It should be emphasized that the procedure is up to now 
independent on the realization of the basic vectors. As shown by Doi \cite{doi} the 
average of an arbitrary physical quantity $\mathcal{B}(\vec n)$ can be calculated by the average of the 
corresponding operator $B(t)$
\begin{equation}
\langle B(t) \rangle = \sum_{n_i} P(\vec n,t) \mathcal{B}(\vec n) = 
\langle s \mid B \mid F(t) \rangle 
\label{fo3}
\end{equation} 
with the state function $\langle s \mid = \sum \langle \vec n \mid$. Defining the density 
operator $\rho = \mid F(t) \rangle \langle s \mid$ the mean value can be even expressed in 
the conventional manner as 
$$
\langle B(t) \rangle = {\rm Tr} \left(\rho B(t)\right)\,. 
$$
The evolution equation for an operator 
$B(t)$ reads now  
\begin{equation}
\partial_t \langle B \rangle = \langle s \mid [B(t),L]_- \mid F(t) \rangle
\label{kin}
\end{equation}
As the result of the procedure, all the dynamical equations governing the 
classical problem are determined by the structure of the evolution operator 
$L$ and the commutation rules of the operators. In our case the dynamics will 
be realized by spin-flip processes.

\section{Coupling to Heat Baths}

\noindent Introducing Pauli-operators satisfying the commutation relation
\begin{equation}
[d_i, d^{\dagger}_j ] = \delta _{ij} (1 - 2 d_i d^{\dagger}_i)\,,
\label{com}
\end{equation} 
the evolution operator of a flip-process at lattice site $i$ reads \cite{sctr}, compare also \cite{gs} 
\begin{equation}
L_i = \lambda (d^{\dagger}_i - d_i d^{\dagger}_i) + \gamma (d_i - d^{\dagger}_i d_i)\,.
\label{fli1}
\end{equation}
Here the flip-rates $\lambda$ and $\gamma$ are parameters which are temperature dependent in case 
the system is coupled to a heat bath. As demonstrated in \cite{sctr,st1,mts} such a coupling can be 
directly incorporated into the quantum formulation by replacing the operator in Eq.~(\ref{fli1}) through 
\begin{equation}
L = \nu \sum_i \left[ (1 - d^{\dagger}_i) \exp(-\beta H/2) d_i \exp(\beta H/2)+ 
(1 - d_i) \exp(-\beta H/2) d^{\dagger}_i \exp(\beta H/2) \right]\,. 
\label{fli4}
\end{equation}
The remaining parameter $\nu$ is determined by the microscopic time scale given by the duration of a 
single spin-flip. The quantity $\beta = T^{-1}$ is the inverse temperature (in units of $k_B$) of the heat 
bath and $H$ describes the static interaction. For a further motivation of this approach and the relation 
to the Glauber model see also see \cite{mts}. Now let us generalize the model by including two heat baths with 
different temperatures $T$ and $T^{\prime}$. A possible extension of Eq.~(\ref{fli4}) is given by  
\begin{equation}
L= \nu \sum_i \left[ (1-d^{\dagger}_i) e^{-H_i/2T'} d_i e^{H_i/2T} + (1 - d_i) e^{-H_i/2T'} 
d^{\dagger}_i e^{H_i/2T} \right]
\label{fli2}
\end{equation}
The two reservoirs are coupled directly to each lattice point $i$, therefore the Hamiltonian is not the 
global one but a local energy functional. To illustrate the approach let us 
discuss the simplest case where the Hamiltonian is given by 
\begin{equation}
H_i = (\varepsilon _i - \mu ) d^{\dagger}_i d_i\,.
\label{ham}
\end{equation}
Here $\varepsilon $ is a characteristic energy and $\mu $ is the chemical potential. 
Using the algebraic properties of the Pauli-operators we get
\begin{equation}
L = \nu \sum_i \left[ (1 - d^{\dagger}_i)d_i \exp((\varepsilon _i- \mu )/2T ) + 
(1 - d_i) d^{\dagger}_i \exp(-(\varepsilon - \mu )/"T^{\prime} \right]\,. 
\label{fli3}
\end{equation}
Instead of the lattice gas variable $n_i = d^{\dagger}_i d_i$ with the eigenvalues $0,\, 1$ we can introduce 
a spin variable by $\sigma _i = 1 - 2 n_i$. Thus the empty state $\mid 0\, \rangle$ corresponds to the 
spin-up state $\mid \uparrow \rangle$ and the occupied state $\mid 1\, \rangle$ is related to the spin-down 
state $\mid \downarrow \rangle$. Having regard to Eq.~(\ref{ham}) we find the non-zero terms of the 
evolution operator (\ref{fli2})   
\begin{eqnarray}
\exp(- H_i/2T^{\prime}) d_i \exp ( H_i/2T ) \mid 1 \rangle &=& \exp(\varepsilon _i - \mu )/2T \mid 0 \rangle \nonumber\\
\exp(- H_i/2T^{\prime}) d^{\dagger}_i \exp ( H_i/2T ) \mid 0 \rangle &=& \exp(-(\varepsilon _i - \mu )/2T^{\prime}) 
\mid 1 \rangle
\label{fli5}
\end{eqnarray} 
The flip process $\mid \downarrow \rangle \to \mid \uparrow \rangle $ is triggered by the heat bath at 
temperature $T$ whereas the inverse process $\mid \uparrow  \rangle \to \mid \downarrow  \rangle $ is 
activated by the bath at $T^{\prime}$. Here we have assumed that the activation energy $\varepsilon $ is 
the same for bath baths. However a generalization to different activation energy and consequently different 
chemical potential is possible. Remark that there is no further 
restrictions for the chemical potential $\mu $
Using Eq.~(\ref{kin}) and the algebraic properties of 
Pauli--operators, the evolution equation for the averaged density reads
\begin{equation}
\nu^{-1} \partial_t \langle n_i \rangle = 
\exp(-(\varepsilon _i - \mu )/2T')\, \langle 1 - n_i \rangle - 
\exp((\varepsilon _i - \mu)/2T)\, \langle n_i \rangle  
\label{e1}
\end{equation}
This equation can be solved exactly and exhibits a stationary solution of the form 
\begin{equation}
\langle n_i \rangle_{s} = \frac{1}{\exp((\varepsilon _i - \mu)/T_e)+ 1} \quad\mbox{with}\quad
\frac{1}{T_e} = \frac{1}{2} \left[\frac{1}{T} + \frac{1}{T'} \right]
\label{e1a}
\end{equation}
Obviously, the effective temperature is not the mixing temperature of both baths. 
In a spin representation we obtain
\begin{equation}
\langle \sigma _i \rangle_s = \frac{e^{(\varepsilon _i- \mu)/2T} -  e^{-(\varepsilon _i -\mu)/2T'}}
{e^{(\varepsilon _i - \mu)/2T} + e^{-(\varepsilon - \mu)/2T'}} 
\label{e2}
\end{equation}
In the special case $T = T'$ the stationary solution coincides with the 
conventional equilibrium solution
\begin{equation}
\langle \sigma _i \rangle_s = \tanh\frac{\varepsilon _i - \mu}{2T}
\label{eq}
\end{equation}
If the temperature of one of the heat baths tends to infinity (for instance 
$T' \to \infty$) the stationary solution is 
$$
\langle \sigma \rangle_s = \tanh\frac{\varepsilon_i - \mu}{4T}
$$
When both temperatures $T$ and $T'$ are infinitesimal different from each other 
$T' = T + \triangle T$  
the averaged spin is
\begin{equation}
\langle \sigma \rangle_s = \tanh\frac{\varepsilon_i - \mu }{2T}
\left[ 1 - \frac{\triangle T (\varepsilon_i - \mu )}{2 T^2 \sinh (\varepsilon_i - \mu )/ T}\right]
\label{eq1}
\end{equation}
\noindent The relaxation time $t_r$ related to the Eq.(\ref{e1}) is simply given by 
\begin{equation}
(\nu t_{\rm rel})^{-1} = \exp(\frac{\varepsilon_i - \mu}{2T}) + \exp(-\frac{\varepsilon_i - \mu}{2T'})
\end{equation}
The relaxation time for $T' \ne T´$ is either enhanced for $T' < T$ or 
diminished in the opposite case. In particular this behavior can be observed for 
small difference between both baths resulting in  
$$
(\nu t_{\rm rel})^{-1} = 2 \cosh\left(\frac{\varepsilon_i - \mu }{2T}\right) + 
\frac{\triangle T \varepsilon_i }{2 T^2}e^{-(\varepsilon_i - \mu )/2T}
$$

\section{Phase Transition}

\noindent Now let us study Eq.~(\ref{e1}) for a magnetic system, where the energy $\varepsilon_i$ at the lattice 
site $i$ depends on the surrounding spin configuration. The chemical potential is in that case zero and the 
interaction is assumed to follow the Ising type. In the simplest mean field approximation the energy is 
given by 
\begin{equation}
\varepsilon_i = 2\, [\,h_i + \sum_j J_{ij} \langle \sigma_j \rangle\, ]\, , 
\label{le0}
\end{equation}
where $h_i$ is an external field and $J_{ij}$ is the interaction between the $z$ nearest neighbors. Firstly 
the homogeneous case in zero field is discussed. Thus, we have  
\begin{equation}
\varepsilon = 2 J z \langle \sigma \rangle \equiv 2 T_0 \langle \sigma \rangle \,,
\label{le}
\end{equation}
whereby $T_0$ plays the role of the critical temperature for the conventional case with only one heat bath. 
Inserting Eq.~(\ref{le}) in Eq.~(\ref{e1}) we get 
\begin{equation} 
\nu ^{-1} \partial_t \langle \sigma \rangle = \exp\frac{T_0 \langle \sigma \rangle}{T} - 
\exp (-\frac{T_0 \langle \sigma \rangle}{T^{\prime}})  - 
\langle \sigma \rangle \left[ \exp \frac{T_0 \langle \sigma \rangle}{T} + 
\exp (-\frac{T_0 \langle \sigma \rangle}{T^{\prime}}) \right]\,.   
\label{le1}
\end{equation}
According to the flip rules defined in Eq.~(\ref{fli5}), this equation is invariant against the simultaneous 
symmetry transformation $\sigma \longleftrightarrow - \sigma $ and $ T \longleftrightarrow T^{\prime}$. 
Thus, making an expansion with respect $\langle \sigma \rangle $ we find 
\begin{eqnarray}
\nu ^{-1} \partial_t \langle \sigma \rangle &=& - r \langle \sigma \rangle  
+ b \langle \sigma \rangle^2 - u \langle \sigma \rangle^3,\nonumber\\
{\rm with}\quad r &=& 2T_0 \left[\frac{1}{T_e} - \frac{1}{T_0} \right],\quad 
b = T_0^2 \left[\frac{1}{T} - \frac{1}{T^{\prime}}\right]\left[\frac{1}{T_e} - \frac{1}{T_0}\right],\nonumber\\
u &=& \frac{T_0^{2}}{2}\left[ \frac{1}{T^2} + \frac{1}{T'^2}\right] - \frac{T_0^3}{6} 
\left[\frac{1}{T^3} - \frac{1}{T'^3}\right]\,.
\label{le2}
\end{eqnarray}
One can easily check that for $T = T^{\prime}$ the conventional Landau expansion results. 
Otherwise for two different heat baths with $ b \not= 0$ the extended symmetry allows a quadratic 
term in $\langle \sigma \rangle$. Thus one concludes that a first order transition is dynamically 
induced. To illustrate the situation in more detail let us derive directly from Eq.~(\ref{le2})
an analog of the free energy for a two bath system:
\begin{eqnarray}
\nu ^{-1} \frac{\partial \langle \sigma \rangle}{\partial t} &=& 
- \frac{\partial F}{\partial \langle \sigma \rangle}\quad\rm{with}\nonumber\\ 
F &=& F_0 + \frac{r}{2} \langle \sigma \rangle ^2 - \frac{b}{3}\langle \sigma \rangle ^3 + \frac{u}{4} 
\langle \sigma \rangle ^4 
\label{le3}
\end{eqnarray}
From here we find another peculiarity of the two bath system, namely the stationary solution is different 
from the minimum of the free energy. The stationary solution follows from Eq.~(\ref{e2}) with Eq.~(\ref{le}) 
to be 
\begin{equation}
\langle \sigma \rangle_s = \tanh \frac{T_0 \langle \sigma \rangle_s}{T_e}
\label{stat}
\end{equation}
Apparently this solution is totally different from the minimum of the free energy, which follows 
from Eq.~(\ref{le2}) by setting $\partial_t \langle \sigma \rangle = 0$. 

\noindent Now let us consider the case that $T^{\prime} = T + \triangle T$. The coefficient, defined in Eq.~(\ref{le2}), 
are for temperatures in the vicinity of $T_0$ 
\begin{equation}
r = 2 [ \tau + \frac{\triangle T}{2T_0}],\quad b = - \tau \frac{\triangle T}{T_0},\quad 
u = \frac{2}{3}[ 1 - \frac{3 \triangle T}{4 T_0} ]\quad {\rm with}\quad \tau = \frac{T - T_0}{T_0} 
\label{le4}
\end{equation}
When $\triangle T \ll T \simeq T_0$ the coefficient b can be neglected leading to a stationary solution 
$$
\langle \sigma \rangle_s = \pm \sqrt{\frac{3}{T_0} \left[ (T_0 - \frac{\triangle T}{2}) - T \right]}\,.
$$
In that case a second order phase transition results where the critical temperature is shifted to 
$T_0 - \frac{\triangle T}{2}$.

\noindent In case of an inhomogeneous field $h_i$ we calculate the response function defined by
\begin{equation}
\chi _{ij} = \left.\frac{\partial \langle \sigma _i \rangle }{\partial h_j}\right|_{h_j = 0}\,.
\label{sus}
\end{equation}
Inserting Eq.~(\ref{le0}) into Eq.~(\ref{e1}), then the 
response function fulfills after performing Fourier transformation the following equation
\begin{eqnarray}
\partial_t \chi (\vec q, t) &=& - R(\vec q) \left[ \chi (\vec q, t) - \frac{\Pi (T,\,T')}{ R(\vec q)}\right]\quad 
\rm{with}\nonumber\\
R(\vec q) &=& \frac{e^{T_0\langle \sigma \rangle _s/T} + e^{-T_0\langle \sigma \rangle _s/T^{\prime}} - 
J(\vec q ) \Pi (T,\,T^{\prime})}{\Pi (T,\,T^{\prime})}\nonumber\\
\Pi (T,\,T') &=& \frac{e^{T_0 \langle \sigma \rangle_s/T}(1 - \langle \sigma \rangle _s)}{T} + 
\frac{e^{-T_0 \langle \sigma \rangle _s/T'}(1 + \langle \sigma \rangle_s)}{T^{\prime}}\,. 
\label{sus1}
\end{eqnarray}
Here $\langle \sigma \rangle (t)$ is replaced by the stationary magnetization $\langle \sigma \rangle_s$, which 
satisfies Eq.(\ref{stat}). In case of long wave excitations it is convenient to expand the interaction as
$$
J(\vec q) = J(0) (1 - c \vec q\,^2)\quad \rm{with}\quad J(0) = J z = T_0\,.
$$
Inserting this expression into the solution of Eq.~(\ref{sus1}) we get in the limit $t \to \infty$ the  
stationary susceptibility of the form
\begin{eqnarray}
\chi_s^{-1} (\vec q\,) &=& c T_0\, \vec q\,^2 + r \nonumber\\ 
\rm{with}\quad r &=& \frac{T_e - T_0\,(1 - \sigma_s\,^2\,)}{1 - \sigma_s\,^2}\,.
\label{sus2}
\end{eqnarray}
In the special case $T = T^{\prime}$ it results $r = [T - T_0( 1 - \sigma_s^2)]/(1 - \sigma_s^2) \simeq T - T_0 $ 
as expected.

\section{Conclusions}

\noindent We have considered spin-flip processes where the rates are conditioned by two different 
heat baths. The situation in mind can be analyzed in a seemingly compact form using the master 
equation in a quantum Hamilton formalism. Using this formalism we find an evolution 
equation for the averaged occupation number, where the stationary solution gives rise to 
a generalized Fermi-Dirac distribution with an effective temperature. This temperature is not the mean value 
of both baths. In terms of equilibrium statistics such a distribution is obtained by coupling separate 
baths to each energy level. Instead of using a lattice gas variable we rewrite the evolution equation in terms 
of spin variables and end up with a Landau-like expansion for the order parameter. From here we conclude the 
existence of a free energy functional, where the coefficients depends on both temperatures. Owing to 
the two reservoirs the system allows a new symmetry consisting of the invariance of the evolution 
equation against the change of the spin orientation and simultaneously the interchange of the baths. 
Consequently, a first order phase transition is dynamically induced. In a further step both baths should 
be coupled leading to a temperature flow.


\newpage
\begin{thebibliography}{90}

\bibitem{gard} C. W. Gardiner, {\em Handbook of Stochastic Methods for Physics, Chemistry and
Natural Science} (Springer-Verlag, Berlin, 2004). 

\bibitem{vk} N. G. van Kampen, {\em Stochastic Processes in Physics and Chemistry} (North-Holland, 
Amsterdam, 1992). 

\bibitem{sun} S. X. Sun, Phys. Rev. Lett. {\bf 96}, 210602 (2006).

\bibitem{doi} M. Doi, J.Phys.A: Math. Gen. {\bf 9} 1465, 1479 (1976).

\bibitem{gra} P. Grassberger and M. Scheunert Fortschr. Physik {\bf 28}, 547 (1980).

\bibitem{sp} H. Spohn {\em Large Scale Dynamics of Interacting Particles} 
(New York: Springer, 1991).

\bibitem{satr} S. Sandow and S. Trimper, Europhys. Lett., {\bf 21}, 799 (1993).

\bibitem{sti} R. B. Stinchcombe, Physica A {\bf 224}, 248 (1996).

\bibitem{mg} D. C. Mattis and M. L. Glasser, Rev. Mod. Phys. {\bf 70}, 979 (1998).

\bibitem{gs} G. M. Sch\"utz, in {\em Phase Transitions and Critical Phenomena} edited 
by C. Domb and L. Lebowitz (Academic Press, London, 2001), Vol.19\,.

\bibitem{sctr} M. Schulz and S. Trimper, Phys. Rev. B {\bf 53}, 8421 (1996).

\bibitem{st1} M. Schulz and S. Trimper, Phys. Lett. A {\bf 227}, 172 (1997). 

\bibitem{mts} T. Michael, S. Trimper, and M. Schulz, 
to be submitted, Phys. Rev. E.

\bibitem{cglv} Z. Cheng, P. L. Garrido, J. L. Lebowitz, and J. L. Vall\'{e}s, 
Europhys. Lett. {\bf 14}, 507 (1991).

\bibitem{tag} P. Tamayo, F. J. Alexander, and R. Gupta, Phys. Rev. E {\bf 50}, 3474 (1994).

\bibitem{br} K. E. Bassler and Z. R\'{a}cz, Phys. Rev. Lett. {\bf 73}, 1320 (1994).

\bibitem{e} P. Espa\~{n}ol, Europhys. Lett. {\bf 40}, 631 (1997).

\bibitem{s} A. Szolnoki, J. Phys. A: Math. Gen. {\bf 30}, 7791 (1997). 

\bibitem{vrmh} S. Velasco, J. M. M. Roco, A. Medina, and A. C. Hern\'{a}ndez, 
Phys. Rev. Lett {\bf 78}, 3241 (1997).

\bibitem{sb} I. M. Sokolov and A. Blumen, J. Phys.A: Math.Gen. {\bf 30}, 3021 (1997).

\bibitem{vs} D. P. Visco, Jr and S. Sen, Phys. Rev. E {\bf 58}, 1419 (1998)

\bibitem{rws} R. S. Reid, W. C. Ward, and G. W. Swift, Phys. Rev. Lett. {\bf 80}, 4617 (1998).

\bibitem{pel} R. Exartier, L. Peliti, Phys. Lett. A {\bf 261}, 94 (1999).

\bibitem{s1} A. Szolnoki, Phys. Rev. E {\bf 60}, 2425 (1999).  

\bibitem{ta} S. Trimper, S. Arzt, Int. J. Mod. Phys. {\bf 13}, 375 (1999).

\bibitem{raf} O. M. Ritter, P. C. T. D'Ajello, and W. Figueiredo, Phys. Rev. E {\bf 69}, 016119 (2004).

\bibitem{b1} C. VandenBroeck, Phys. Rev. Lett. {\bf 95}, 190602 (2005).

\bibitem{zb} X.-P. Zhang and J.-D. Bao, Phys. Rev. E {\bf 73}, 061103 (2006).

\bibitem{kn} T. S. Komatsu and N. Nakagawa, Phys. Rev. E {\bf 73}, 065107(R) (2006).

\bibitem{ah} D. Ben-Amotz and J. M. Honig, Phys. Rev. Lett. {\bf 96}, 020602 (2006).

\bibitem{a} P. Attard, J. Chem. Phys. {\bf 124}, 224103 (2006).

\bibitem{hh} P. C. Hohenberg and B. I. Halperin, Rev. Mod. Phys. {\bf 49}, 435 (1977).

\end{thebibliography}
\end{document}